# Identification of microRNA clusters cooperatively acting on Epithelial to Mesenchymal Transition in Triple Negative Breast Cancer

Cantini L[1,2,3,4]#, Bertoli G[5]#, Cava C[5], Dubois T[1,2,6], Zinovyev A[1,2,3,4], Caselle M[7], Castiglioni I[5]*, Barillot E[1,2,3,4], Martignetti L[1,2,3,4]*

1 Institut Curie, 26 rue d'Ulm, F-75005 Paris, France
2 PSL Research University, F-75005 Paris, France
3 Inserm, U900, F-75005, Paris France
4 Mines Paris Tech, F-77305 cedex Fontainebleau, France
5 Institute of Molecular Bioimaging and Physiology, National Research Council (IBFM-CNR) Italy
6 Institut Curie, PSL Research University, Department of Translational Research, Breast Cancer Biology Group - Paris, France
7 Department of Physics and INFN, Università degli Studi di Torino - Turin, Italy

# Contributed equally
* Corresponding Authors

## Abstract

MicroRNAs play important roles in many biological processes. Their aberrant expression can have oncogenic or tumor suppressor function directly participating to carcinogenesis, malignant transformation, invasiveness and metastasis. Indeed, miRNA profiles can distinguish not only between normal and cancerous tissue but they can also successfully classify different subtypes of a particular cancer.

Here, we focus on a particular class of transcripts encoding polycistronic miRNA genes that yields multiple miRNA components. We describe clustered MiRNA Master Regulator Analysis (ClustMMRA), a fully redesigned release of the MMRA computational pipeline (MiRNA Master Regulator Analysis), developed to search for clustered miRNAs potentially driving cancer molecular subtyping. Genomically clustered miRNAs are frequently co-expressed to target different components of pro-tumorigenic signalling pathways. By applying ClustMMRA to breast cancer patient data, we identified key miRNA clusters driving the phenotype of different tumor subgroups. The pipeline was applied to two independent breast cancer datasets, providing statistically concordant results between the two analysis. We validated in cell lines the miR-199/miR-214 as a novel cluster of miRNAs promoting the triple negative subtype phenotype through its control of proliferation and EMT.

# 1. Introduction

MicroRNAs (miRNAs) are small RNA molecules emerged as important regulators of gene expression at the post-transcriptional level. They have been shown to be involved in the regulation of all essential functions of the cells from differentiation and proliferation to apoptosis[1]. Each miRNA possesses hundreds of target genes, and a single gene can be targeted by several miRNAs[2], giving rise to complex interaction networks, currrently very partially characterized.

Multiple studies demonstrated the importance of miRNAs in all the cancer hallmarks defined by Hanahan and Weinberg[3] and indicated that they might function as oncogenes or tumor suppressors[4–7]. Further experimental evidences suggested that specific miRNAs may also have a role beyond the cancer onset and directly participate in cancer invasiveness and metastasis[6,8]. Indeed, miRNA profiles can distinguish not only between normal and cancerous tissue but they can also successfully classify different subtypes of a particular cancer[9,10], notably of breast cancer[11–13].

In this work, we focused our attention on a particular class of transcripts encoding polycistronic miRNA genes that yields multiple miRNA components. A famous example of this class of transcripts is the mir-17/92 polycistronic oncogene that plays a role in the development of various cancer types, especially in their most aggressive form[14]. Genomically clustered miRNAs of mir-17/92 are simultaneously expressed and target different components of the signaling cascade as well as the downstream effectors of pro-tumorigenic signalling pathways[15–17]. Deep sequencing of triple negative breast cancer (TNBC) samples revealed a threefold increase of miR-17/92 levels[12]. Other studies in breast cancer have shown that mir-106b/25 cluster activates TGF-β signaling and epithelial-mesenchymal transition (EMT)[18] and miR-221/222 cluster is a key regulator of luminal breast cancer tumor progression[19].

Since more than 30% of annotated human miRNAs are organized in genomic clusters, we can expect to find other oncogenic / tumour suppressor polycistronic miRNAs that are co-expressed to jointly regulate molecular pathways involved in cancer malignancy. Existing computational approaches for the identification of master miRNA regulators involved in cancer onset and subtyping are typically designed to detect the effect of a single miRNA (see review in[20]). However, miRNAs have been shown to frequently act in a combined manner, jointly regulating proteins in close proximity of the protein-protein interaction network[21] and functionally related genes[22–25]. The underlying assumption of this work is that this mode of action might be true also for genomically clustered miRNAs. Indeed, it has already been shown that clustered miRNAs carry out pervasive cotargeting[26].

Here we present Clustered MiRNA Master Regulator Analysis (ClustMMRA), a fully redesigned release of the MiRNA Master Regulator Analysis (MMRA)[25,26] pipeline, developed to search for clustered miRNAs potentially driving cancer subtyping. MMRA was designed for miRNA underlying tumor subtypes, a comparison characterized by much lower variation than cancer versus normal conditions. The results of the MMRA pipeline were experimentally validated, proposing a set of four miRNAs predicted to drive the stem-like aggressive colorectal cancer subtype[27].

ClustMMRA extends MMRA to a model in which multiple miRNAs belonging to the same genomic cluster coordinately target functionally related genes driving the phenotype of a particular cancer subtype. As the MMRA pipeline, ClustMMRA is a multi-step workflow that requires in input miRNA/mRNA expression profiles from matched tumor samples classified in different subtypes according to subtype-specific gene signatures. The final output of ClustMMRA provides key miRNA clusters contributing to the regulation of particular subtypes of the disease.

We tested this novel pipeline to search for oncogenic / tumour suppressor polycistronic miRNAs driving breast cancer subtypes. ClustMMRA was applied to two independent breast cancer datasets whose samples were previously classified into four subtypes (luminal A, luminal B, HER2+ and triple negative). We obtained statistically concordant results between the two analysis, identifying five clusters of miRNAs with aberrant expression in a specific subtype of both datasets. Among them, miR-199a/214 on chromosome 1 was found to be down-regulated in the triple negative subtype and associated to EMT regulation. Functional validation in cell lines confirms the regulatory effect of this cluster in shaping the triple negative subtype phenotype through its control of proliferation and EMT. Overall, our computational pipeline and experimental validations characterize a new genomic cluster of miRNAs implicated in the TNBC phenotype that might be further explored in diagnosis and therapeutic strategies. In addition, we evinced a cooperative mechanism for the regulatory activity of genomically clustered miRNAs.

## 2. Results

## 2.1 From single miRNA to clusters of miRNAs: ClustMMRA

The MMRA pipeline is here extended to search for genomically co-clustered miRNAs potentially driving cancer subtyping. Similar to MMRA, the workflow of ClustMMRA (see Figure 1) consists of subsequent filtering steps: (i) differential expression analysis of clustered miRNAs; (ii) target enrichment analysis and (iii) network analysis. While a miRNA cluster is usually transcribed as a single unit[28–32], the expression of mature miRNAs in the same cluster might not be highly correlated due to regulatory events in the maturation processes[28,31].

Clusters of miRNAs are identified based on their genomic organization as reported in Methods. In step (i), the subtype-specific expression of each miRNA is assessed by Kolmogorov-Smirnov (KS) statistical test and fold change cutoff. Clusters having at least two miRNAs with subtype-specific expression change in the same direction (both up-regulated or down-regulated) are selected for step (ii).

In step (ii), we extract miRNA clusters having their predicted targets enriched for the gene signature of the corresponding subtype. Only miRNAs of the cluster classified as differentially expressed in step (i) are considered in step (ii). The targets of individual miRNAs have been predicted using four different databases (miRTarBase 2.5, doRiNA-PicTar 2012, microRNA.org 2010, PITA 2007 and TargetScan 7.1), requiring the prediction by at least two of them. The set of targets of a cluster has been defined as the union of the targets of individual miRNAs. The objective of step (i) and (ii) is to identify co-clustered and co-expressed miRNAs potentially regulating a gene expression signature in a joint manner, without necessarily having a high overlap in terms of target genes[23]. Finally, in step (iii) a miRNA-mRNA interaction network is constructed for each selected cluster using the ARACNE algorithm[33,34]. In this step, we identify modules of co-clustered miRNAs and interacting genes, including indirect interactions, that are believed to participate in the phenotype of a given cancer subtype (we call these modules *regulons*). Unlike the results of the MMRA pipeline, in which *regulons* can include only one miRNA, the ones identified by the ClustMMRA pipeline contain multiple miRNAs of the genomic cluster. Interference of indirect interactions may introduce links between miRNAs and spurious genes in the *regulon*. A Fisher's exact test has been performed to evaluate the statistical significance of the overlap between the genes included in each *regulon* and the gene signature of the associated subtype.

## 2.2 Identification of regulatory miRNA clusters underlying breast cancer subtypes

We applied ClustMMRA to identify polycistronic miRNAs underlying breast cancer molecular subtypes. For this study, two independent datasets were used, a first paired miRNA/mRNA expression dataset from a in-house cohort of 129 breast carcinoma tumour samples (which we refer to as Curie dataset[35,36] and a second dataset from The Cancer Genome Atlas project composed of 397 samples[37]. In both datasets, individual samples were assigned to four subtypes (luminal A, luminal B, HER2+ and triple negative) based on the immunohistochemical staining of estrogen (ER), progesterone (PR) and HER-2 (ERBB2) receptors.

### 2.2.1 ClustMMRA application to Curie and TCGA datasets

Expression data required for running ClustMMRA were pre-processed as described in Methods and the signatures for breast cancer subtypes were defined using the approach proposed in[38] (see Methods). We applied the ClustMMRA pipeline on Curie and TCGA datasets separately. In the first step, genomically co-clustered miRNAs having a subtype-specific expression were identified. In this step, 28 and 47 out of 131 analyzed clustered miRNAs were selected for Curie and TCGA datasets, respectively (see Supplementary Table S1). Of these, 18 clusters were in common between the two datasets (p-value<7e-04), revealing a significantly concordant expression pattern of co-clustered miRNAs. Among these co-clustered and co-expressed miRNAs, some are differentially expressed in multiple subtypes (18 and 37 clusters for Curie and TCGA respectively), with 15 out of 18 and 21 out of 37 differentially expressed in basal-like and luminal A with opposite sign.

In step (ii), 10 out of 28 (Curie) and 16 out of 47 (TCGA) subtype-specific miRNA clusters were found to have their predicted targets enriched in genes belonging to the corresponding gene signature. The output of step (ii) (see Supplementary Table S2) has an intersection of 7 elements between the two datasets (p-value <1e-05). In the step (iii) of ClustMMRA, a *regulon* for each miRNA cluster selected in step (ii) was constructed. The *regulons* were tested for enrichment in gene signature. 7 out of 10 and 9 out of 16 clusters passed this last selection step in Curie and TCGA datasets, respectively. These clusters constitute the final output of ClustMMRA and are reported in Table 1. After this last step, the output in common between the two datasets contains 5 clusters (p-value <8e-06). The significant overlap between results obtained from the analysis of two independent datasets with ClustMMRA supports the reliability of this approach. Notably, the results have an intersection with increasing statistical significance at each step of the pipeline. This trend confirms the accuracy of the proposed pipeline in selecting candidate clusters underlying cancer subtypes.

Some results obtained with ClustMMRA in the breast cancer study have already been validated in the literature. MiR-493/136 and miR-379/656 clusters in the chromosomal region 14q32 have been reported as tumor suppressors in different types of human cancer[39–41], including breast cancer[42]. Silencing of multiple miRNAs encoded in these clusters was shown to increase the proliferation and invasion of ovarian [43], melanoma[44] or oral squamous carcinoma[39] cells. The X-chromosome-located miR-532/502 cluster has been previously associated to cancer. In particular, this was found up-regulated in triple-negative breast cancer cells[45] and the regulatory circuit miR-502/H4K20 methyltransferase SET8 was described as a key regulator of breast cancer pathobiology[46].

**Table 1. Clusters of miRNAs identified by ClustMMRA in breast cancer TCGA and/or Curie datasets.**

| Cluster of miRNAs | Chromosome position | Number of deregulated miRNAs in the cluster | Cluster expression in subtypes | Gene signature expression in subtypes | Dataset results |
|---|---|---|---|---|---|
| **miR-199a/214** | Chr1 | 3 | Down in Basal-like | Up in Basal-like | **Curie and TCGA** |
| **miR-493/136** | Chr14 | 8 | Down in Basal-like | Up in Basal-like | **Curie and TCGA** |
| **miR-379/656** | Chr14 | 42 | Down in Basal-like | Up in Basal-like | **Curie and TCGA** |
| **miR-512/373** | Chr19 | 46 | Up in Basal-like | Up in Basal-like | **Curie and TCGA** |
| **miR-532/502** | ChrX | 8 | Up in Basal-like | Down in Basal-like | **Curie and TCGA** |
| miR-449a/449c | Chr5 | 3 | Down in Basal-like | Down in Basal-like | TCGA |
| miR-653/489 | Chr7 | 2 | Down in Basal-like | Down in Basal-like | TCGA |
| miR-548aa/548d | Chr8 | 2 | Up in Basal-like | Down in Basal-like | TCGA |
| miR-421/374c | ChrX | 3 | Up in Basal-like | Up Basal-like | TCGA |
| miR-99a/let-7c | Chr21 | 2 | Down in Basal-like | Up Basal-like | Curie |
| miR-450b/424 | ChrX | 6 | Down in Basal-like | Up Basal-like | Curie |

**2.2.2 Comparison of ClustMMRA with the pipeline for the identification of single master miRNA regulators (MMRA)**

We compared the results of ClustMMRA in the breast cancer study with those obtained by applying to the same dataset the MMRA pipeline for the identification of single master miRNA regulators. The goal is to investigate if the regulatory effect of a cluster can be detected by studying the effect of individual miRNAs belonging to the same cluster.

We applied MMRA to the Curie dataset, using in each step the same thresholds employed for ClustMMRA. If at least two miRNAs of a given cluster are included in the output of MMRA, we consider this cluster as detected in the single-miRNA pipeline. Interestingly, 4 out of 7 clusters detected by ClustMMRA (miR-199a/214, miR-493/136, miR-512/373 and miR-450b/424) were not detected by MMRA.

This difference between the output of the two pipelines is given by the target enrichment analysis in step (ii) and the network analysis in step (iii). In fact, the 4 clusters missing in the final output of MMRA are included in the output of step (i), since they have at least 2 differentially expressed miRNA genes. They are filtered out in step (ii) since no miRNA gene in these clusters, when analyzed individually, reaches a significant enrichment of signatures genes in its targets for a certain subtype. This observation supports the hypothesis that co-clustered miRNAs participate in regulating the gene expression signature of a given cancer subtype without necessarily having a high overlap in terms of common target genes.

**2.2.3 Prioritization of miRNA clusters for functional validation in cell lines**

Before experimental validation of the ClustMMRA output, prioritization of results was performed. We considered the 5 clusters identified both in TCGA and Curie datasets. For the *regulons* associated to each cluster, the nodes present in both TCGA and Curie datasets were kept, obtaining a network for each *regulon* with size of about 100 nodes. Then, biological processes and pathways associated to these *regulons* were identified through Fisher's exact enrichment test, using MSigDB [47] as reference collection of signatures for pathways and biological functions. The complete list of MSigDB pathways resulting from this analysis (FDR < 0.05) is reported in Supplementary Table S3.

Overall, the network analysis shows a regulation of EMT, stemness and extracellular matrix by clusters miR-493/136, miR-379/656 and miR-199a/214. Cluster miR-

532/502 is predicted to regulate proliferation and the cell cycle transition from G to M phases. All the *regulons* have been found associated to breast cancer specific signatures, with clusters miR-493/136, miR-379/656 and miR-199a/214 sharing 9 of them ("SCHUETZ_BREAST_CANCER_DUCTAL_INVASIVE_UP","FARMER_BREAST_CANCER_CLUSTER_4","TURASHVILI_BREAST_LOBULAR_CARCINOMA_VS_LOBULAR_NORMAL_DN","CHARAFE_BREAST_CANCER_LUMINAL_VS_MESENCHYMAL_DN","LANDIS_BREAST_CANCER_PROGRESSION_DN","LANDIS_ERBB2_BREAST_TUMORS_324_DN","LIEN_BREAST_CARCINOMA_METAPLASTIC","TURASHVILI_BREAST_DUCTAL_CARCINOMA_VS_DUCTAL_NORMAL_UP","TURASHVILI_BREAST_LOBULAR_CARCINOMA_VS_DUCTAL_NORMAL_UP","TURASHVILI_BREAST_LOBULAR_CARCINOMA_VS_LOBULAR_NORMAL_DN"). Invasive and mesenchymal state signatures confirm the association of these clusters to the basal-like subtype. Other general processes were found enriched in the regulons of these clusters: EMT (including the "HALLMARK_EPITHELIAL_MESENCHYMAL_TRANSITION" signature and multiple GO terms related to the extracellular matrix), stemness ("BOQUEST_STEM_CELL_UP","LIM_MAMMARY_STEM_CELL_UP","IZADPANAH_STEM_CELL_ADIPOSE_VS_BONE_DN" signatures), cell cycle ("IGLESIAS_E2F_TARGETS_UP") and angiogenesis ("GO_VASCULATURE_DEVELOPMENT","GO_CIRCULATORY_SYSTEM_DEVELOPMENT"). Finally, the regulon of cluster miR-532/502 was found enriched in some breast cancer specific signatures clearly linking it to the basal-like subtype ("SOTIRIOU_BREAST_CANCER_GRADE_1_VS_3_UP","FARMER_BREAST_CANCER_BASAL_VS_LULMINAL" and "POOLA_INVASIVE_BREAST_CANCER_UP"). Also, it was observed to be strongly associated to proliferation signatures (e.g. "ZHOU_CELL_CYCLE_GENES_IN_IR_RESPONSE_24HR","GO_MITOTIC_NUCLEAR_DIVISION","GO_MITOTIC_CELL_CYCLE","GO_CHROMOSOME_SEGREGATION","GO_CELL_DIVISION","GO_CELL_CYCLE_PROCESS","CHANG_CYCLING_GENES").

We focused on EMT regulation by miR-199a/214 as an interesting phenotype to validate in basal-like subtype. MiR-199a/214 is the smallest cluster that controls EMT, in terms of miRNA genes. Considering the technical difficulty in producing the over-expression of multiple miRNAs in cell lines, this was chosen as the best candidate to study the combinatorial regulation by co-clustered miRNAs.

### 2.2.4 MiR-199a/miR-214 cluster is underexpressed in TNBC cells

Human miR-199a/miR-214 cluster is encoded by a large non-coding RNA on chromosome 1q24 which produces three mature miRNAs (hsa-miR-199a-5p, hsa-miR-199a-3p and hsa-miR-214). First, we examined by quantitative RT-PCR the expression of the individual mature miRNAs belonging to this cluster in T47D and MDA-MB-231 cells, which are luminal A and TNBC cells respectively[48]. Results show that the three mature miRNAs encoded by the miR-199a/miR-214 cluster are significantly underexpressed in MDA-MB-231 compared to T47D cells (Fig.3).

### 2.2.5 Upregulation of miR-199a/miR-214 cluster decreases TNBC cell proliferation

To test whether the deregulation of miR-199a/miR-214 cluster was sufficient to impact TNBC cells phenotype, MDA-MB-231 cells were treated with sense (S) oligonucleotides encoding for all the three miRNAs of the cluster (miR-214, miR-199a-5p, miR-199-3p) or scramble negative controls. We checked the overexpression of each miRNA of the cluster after transfection by RT-PCR analysis, shown in Fig.4A-C. After confirming the upregulation of single miRNA or all three miRNAs of the cluster in MDA-MB-231, we analyzed the effect of miRNA overexpression on proliferation: individual miRNAs, except miR-199a-3p, and entire miR-199a/miR-214 cluster overexpression reduce the MDA-MB-231 cell number compared to scramble or untreated control (Fig. 5).

### 2.2.6 MiR-199a/miR-214 cluster silencing is associated with EMT-like and invasive phenotype

According to bioinformatic analysis, miR-199a/miR-214 cluster is predicted to modulate EMT genes and cell invasion. To investigate if the expression of this cluster affects the molecular profile of the cells, we analyzed the expression levels of EMT-related genes upon upregulation of a single miRNA of the cluster or the whole cluster through S oligonucleotide treatment. We observed a reduction of EMT marker genes upon both individual miRNAs or entire miR-199a/miR-214 cluster overexpression (Fig.6), as demonstrated by the increase expression of epithelial markers E-cadherin and Beta-catenin and a decrease of the expression level of the mesenchymal marker Slug.

Finally, we used an *in vitro* culture system developed to assess mammary cell propagation in non-adherent, non-differentiated culture conditions and their ability to form discrete clusters of cells termed mammospheres[49]. The ability of the cells to form mammosphere could be considered also a marker of the stemness of the cell population[49]. The formation of such spheroids increases with EMT induction (PMID: 18485877). Our experiments on MDA-MB-231 cells show that the expression of miR-199a/miR-214 cluster is sufficient to compromise mammosphere formation efficiency (Fig.7). In fact, when we overexpressed either miR-214 or miR-199a-5p or miR-199a-3p and the three miRNAs together, we observed a decrease efficacy in mammosphere formation in respect to untreated cells.

## 3. Discussion

Over the last two decades there has been an explosion of research focused on miRNAs involvement in cancer initiation and progression, pointing out the potential of these small RNAs as biomarkers for diagnosis, prognosis and response to treatment. However, the majority of computational and experimental approaches for the identification of master miRNA regulators involved in cancer onset and subtyping are typically designed to detect the regulatory effect of a single miRNA. This can be a limitation in identifying regulation by multiple miRNA species acting cooperatively on cellular pathways and pathological changes.

The computational pipeline here described, ClustMMRA, was specifically designed to search for genomically clustered miRNAs potentially driving cancer subtyping. ClustMMRA provides a computational framework to systematically investigate polycistronic miRNA transcripts involved in cancer subtyping or possibly in other biological contexts. In practice, the use of ClustMMRA can be generalized in order to study other classes of cooperatively acting miRNAs than the case of genomic clusters, such as co-expressed miRNAs from different genomic locations.

In our study, ClustMMRA was applied to search for oncogenic / tumour suppressor polycistronic miRNAs driving breast cancer subtypes, pointing out five novel miRNA clusters whose regulatory effect is potentially associated to the triple negative subtype phenotype. Among them, the miR-199/miR-214 is identified as acting on EMT in TNBC subtype. Our computational and experimental validation of the regulatory effect of miR-199/miR-214 show that the down-regulation of this genomic cluster is associated to appearance of EMT-like phenotype in the TNBC cells. The upregulation of individual miRNAs belonging to the cluster or the entire cluster decreases the expression of a marker of mesenchymal phenotype (i.e., Slug) and increases the expression of epithelial markers (E-cadherin and Beta-catenin). These changes towards an epithelial phenotype, obtained by overexpression on miR-199/miR-214 cluster, diminished the capability of the stem population of MDA-MB-231 lineage of forming mammospheres in suspension. The presence of cancer stem cells has been linked to poor cancer patient survival, as those tumors with a high percentage of cancer stem cells are capable of migrating, invading and colonizing surrounding tissues, surviving in suspension, and creating a secondary tumor[50]. Our results suggest that this cluster of miRNAs is possibly involved in the maintenance of more aggressive phenotype of breast cancer, by controlling the stemness of the population, regulating EMT target genes, and cell proliferation. Finally, our study supports a the hypothesis of miRNA cooperativity from a polycistronic transcript as a possible mechanism of jointly targetting to act on molecular pathways involved in cancer malignancy and subtyping. More accurate measurements and quantitative study might improve the understanding of this cooperative effects.

## 4. Methods

### 4.1 MiRNA cluster annotation

The genomic locations of miRNAs were retreived from miRBase v18[51]. Similar to previous studies[52,53], co-clustered miRNAs are defined as miRNA genes located within 10 Kb of distance on the same chromosome and in the same strand.

### 4.2 Datasets preprocessing

Breast cancer (BRCA) RNA-seq and miRNA-seq Level 3 expression profiles were downloaded from The Cancer Genome Atlas (TCGA) in January 2016. Only those primary tumors profiled for both mRNA and miRNA expression were included in the analysis, obtaining a total of 397 samples. Two expression matrices (one for mRNAs and the second for miRNAs) were normalized obtaining the paired mRNA/miRNA expression dataset here referred to as TCGA. The Curie dataset was generated with microarray technologies (Agilent miRNA microarray kit V3 for miRNAs and Affymetrix U133plus2 for mRNA) and pre-processed following the procedure described in[54].

## 4.2 Definition of a gene signature for each breast cancer subtype

The ClustMMRA pipeline requires as input a gene signature for each disease subtype. Available signatures for breast cancer subtypes, such as the PAM50[55], were not applicable here due to their limited size in terms of number of genes. We thus defined the signatures for our breast cancer study using the approach proposed in[38]. The Curie dataset was used for signature construction, while the TCGA dataset was employed for signature validation. Differential gene expression for each subtype vs. all the other samples was computed by Student's t-test and log fold change cutoff (t-test adjusted p-value < 0.05 and absolute(log fold change) > 0.5). Moreover, to increase the predictive power of the constructed signatures, those genes associated to more than one class according to the previous criteria, or having a difference between the first and second highest absolute(log fold changes) lower than 0.2 were discarded. The choice of thresholds was optimized to maximize the gene association to a unique subtype and the number of genes included in each signature (on average 117 genes per signature). For each subtype, two separated signatures were defined ("down" and "up"), based on the sign of the expression change of their genes. The signatures constructed in this way are available in Supplementary Table S4. The reliability of these signatures were tested in two ways. First, their classification performances were validated on TCGA data. We classified the TCGA samples using our signatures with the Nearest Template Prediction (NTP) method[56], as done in[57,58]. Only 44 out of 397 (11%) samples resulted to be misclassified. Then, the significance of the intersection between our signatures and publicly available ones was evaluated by a Fisher's exact test. The signatures used for this test were obtained from MSigDB[47] plus a specific one derived from[59]. The proliferation signatures were added to test the basal-like subtype, known to be associated to a strong proliferative signal. Highly significant p-values were obtained for the intersection between our newly defined signatures and previously published ones for the same breast cancer subtypes. The above results confirm the classification performances and reliability of the breast cancer signatures here constructed.

## 4.3 Cell culture and miRNA modulation

For *in vitro* studies, we used two human BC epithelial cell lines: T47D and MDA-MB-231 cells (ICLC-Biologic Bank and Cell Factory, Italy). These cell lines were chosen as they represent a model of luminal A and TNBC cell lines, respectively[48]. Following the manufacturer's recommendation, we maintained the cell lines within a humidified atmosphere containing 5% $CO_2$ at 37 °C in DMEM (for T47D cell line) or advanced DMEM (for MDA-MB-231 cell line) cell culture medium (Gibco, Life Technologies), with 10% fetal bovine serum (FBS), 1% penicillin-streptomycin, 2mM glutamine (all from Lonza, Euroclone). Dulbecco Phosphate-Buffered Saline (D-PBS), trypsin, and all the media additives were obtained by Lonza (Euroclone).

The sense (S) oligonucleotide sequence of each miRNA of the cluster has been designed following the sequences indicated in miRbase database[51]. S oligonucleotides were purchased from Sigma.

To obtain the upregulation of each miRNA, S oligonucleotides, resuspended in water, were added three times a day for 3 days directly to the culture medium of the cells (<50% confluency) at a final concentration of 100nM/day[60]. The cells were collected 24,48 or 72h of treatment and different assays were performed (proliferation, mammosphere formation and real time-PCR analysis of miRNAs and EMT genes).

### 4.3 Proliferation assay

Tumor cell proliferation was assessed by following the protocol described in[61]. Briefly, cells were seeded at a confluency of 80000 cells/w in 24 well plates. The cells were added daily with 100nM final concentration of S miR-214, -199a-3p, -199a-5p. The cells were collected and counted at 24,48 or 72h of treatment. A graphic representation of the cell counts was obtained by plotting the number of the total cells at each time point. Experiments were performed three times in triplicate (n = 9).

### 4.5 Mammospheres preparation

After miRNA treatment cells were collected and seeded in non adherent plastic plates (100 cells/ml) in DMEM:F12 (1:1) added with 1% penicillin-streptomycin, 2mM glutamine, 1% Hepes, 10ng/ml bFGF, 20ng/ml B27, 20ng/ml EGF, as described in[62]. Pictures were taken after 10 days of culture in suspension.

### 4.6 RNA isolation, reverse transcription and RT-PCR analysis

Total RNA was isolated using TRIzol reagent (Life Technologies) following the manufacturer's recommendations. To obtain cDNA from total RNA for gene expression analysis, two micrograms of total RNA were reverse transcribed using oligo dT primers in combination with High Capacity cDNA Reverse Transcription kit (Applied Biosystem), following the manufacturer's protocol.

For miRNA analysis, one microgram of total RNA was reverse transcribed using MystiCq microRNA cDNA synthesis kit (Sigma), following the manufacturer's protocol, in order to reverse transcribe polyA-tailed miRNA into cDNA.

RT-PCR analysis was performed using Power Up Sybr Green Master mix (Applied Biosystem, Life Technologies) in an Eco RT-PCR machine (Illumina). All the primers for human mRNA and miRNA amplification were home-made and are described below (Table xx). miRNA amplification was performed using primers designed on the mature miRNA sequence taken from miRbase v18[51]. HPRT and miR-103-3p were used as an internal control for gene expression and miRNA profile analysis, respectively. Primers used are reported in Supp FileXXX

The relative expression of miRNAs and genes was calculated for both T47D and MDA-MB-231 cell lines with the $2^{(-\Delta\Delta C_T)}$ method[63]. Experiments were performed three times in triplicate ($n$ = 9). A $t$ test was calculated.

## 5. Figures

**Figure 1. Schematic representation of the Clustered microRNA Master Regulator Analysis (ClustMMRA) workflow.** The schema reports the data required as initial input, the four analytical steps with the respective outputs, and the final output of the pipeline.

**Figure 2. Pathways controlled by the deregulated miRNA clusters.** A summary of the main biological functions controlled by the different miRNA clusters is here reported. Y-axis of the radarplot corresponds to the sum of the absolute log(p-value) of all the pathways associated to a given function. A,B,C,D correspond to miR-199a/214, miR-493/136, miR-379/656 and miR-532/502, respectively.

**Fig.3 RT-PCR analysis of miRNA expression in T47D vs MDA-MB-231.**
T47D (in white) and MDA-MB-231 (in grey) were analyzed for the expression of miR-214 (A, p-value<0.011), miR-199a-5p (B, p-value<0.003) and miR-199a-3p (C, p-value<0.03). 2^-DDCt method was used for evaluating the expression level of each miRNA. Average±sd of three independent experiments for each cell line are shown. T-test p-value<0.01(**), <0.05(*).

**Fig.4 Mirna modulation in MDA-MB-231 cells.**
MDA-MB-231 cells were treated for 48 hours with 100nM sense (S) oligonucleotide encoding for miR-214, miR-199a-5p, miR-199a-3p or miRNA cluster, respectively. The expression levels of miR-214 (A), miR-199a-5p (B) and miR-199a-3p (C) were evaluated by RT-PCR analysis comparing miRNA-treated cells vs untreated cells. Average±sd of three independent experiments for each cell line are shown. T-test p-value<0.01(**), <0.05(*).

**Fig.5 In vitro analysis of miRNA modulation effect on MDA-MB-231 cells proliferation.**
MDA-MB-231 cells were treated for 24,48,72 hours (h) with sense (S) oligonucleotide encoding for miRNA cluster or single miRNA (miR-214, miR-199a-5p, miR-199a-3p) or a scramble miRNA. The effect of miRNA modulation on cell proliferation is shown. Average±sd of three independent experiments for each cell line are shown. T-test p-value<0.001(***),<0.01(**), <0.05(*).

**Fig.6 Effect of miRNA modulation on EMT marker genes.**
MiRNA modulated MDA-MB-231 cells were used for RT-PCR analysis of EMT marker genes. RT-PCR analysis shows the effect of single miRNA or miRNA cluster modulation vs scramble oligonucleotide treated cells on E-cadherin (A), Beta-catenin (B) and Slug (C). Average±sd of three independent experiments for each cell line are shown. T-test p-value<0.01(**), <0.05(*).

**Fig.7 Effect of miRNA modulation on mammosphere (MM) formation ability.**
MiRNA-modulated MDA-MB-231 were used for MM assay. Pictures of miRNA cluster-treated vs scramble oligonucleotide-treated cells were taken after 10 days of MM formation.

## 7. Acknowledgments

This work was partly supported by ITMO Cancer within the framework of the Plan Cancer 2014–2019 and convention Biologie des Systèmes N°BIO2015–01 (M5 project).

## 8. Authors' contribution

Conceptualization, LC and LM; Data acquisition: LC, TD, GB and IC; Methodology, LC, GB, CC, IC, MC and LM; Validation, GB, CC, IC; Resources, EB and IC; Supervision, EB, IC and LM; Writing – Original Draft, LC, GB and LM; Writing – Review & Editing, LC, GB, CC, IC, TD, MC, AZ, EB and LM.

## 9. Supplementary files

**Supp. File S1.** Supplementary Tables S1-S4